\documentclass[12pt]{article}

\usepackage{graphicx} 

\def\beq{\begin{equation}}
\def\eeq{\end{equation}}
\def\barr{\begin{eqnarray}}
\def\beqa{\begin{eqnarray}}
\def\earr{\end{eqnarray}}
\def\eeqa{\end{eqnarray}}
\def\winf{W_{1+\infty}\ }
\def\u1{\widehat{U(1)}}
\def\v{V\,}
\def\w{W\,}
\def\vb{{\overline V}\,}
\def\wb{{\overline W}\,}

\newcommand{\nl}{\nonumber \\}

\begin{document}

\begin{titlepage}

\begin{center}
\hfill  \quad  \\
\vskip 0.5 cm
{\Large \bf Bosonic Bogoliubov transformations as Lorentz boosts in $(c,{\overline c})=(1,1)$ conformal field theories
with marginal $J{\overline J }$ deformations}

\vspace{0.5cm}

Federico~L.~ BOTTESI$^a$ ,\ \ Guillermo~R.~ZEMBA$^{b,c,}$\\
\medskip
{\em $^a$Facultad de Ingenier\'ia, Universidad de Buenos Aires,}\\
{\em  Av. Paseo Col\'on 850,(C1063ACL) Buenos Aires, Argentina}\\
\medskip
{\em $^b$Facultad de Ingenier\'ia y Ciencias Agrarias,  Pontificia Universidad Cat\'olica Argentina,}\\
{\em  Av. Alicia Moreau de Justo 1500,(C1107AAZ) Buenos Aires, Argentina}\\
\medskip
{\em $^c$Departamento de F\'{\i}sica Te\'orica de Interacciones Fundamentales y Sistemas Complejos, Laboratorio Tandar,}\\
{\em  Comisi\'on Nacional de Energ\'{\i}a At\'omica,} \\
{\em Av.Libertador 8250,(C1429BNP) Buenos Aires, Argentina}\\
\medskip

\end{center}
\vspace{.3cm}
\begin{abstract}
\noindent
We consider conformal field theories with central charges $(c,{\overline c})=(1,1)$ 
that are invariant under the exchange of the holomorphic and antiholomorphic 
sectors, in both bosonic and fermionic realizations that are 
meaningful for condensed matter systems. 
The effect of marginal current-current $(J,{\overline J})$ perturbations 
is to induce a deformation of the Hilbert space given by 
a Lorentz boost in the 2D space of currents, which is identified with 
a Bogoliubov transformation. The rapidity of the boost is determined
by the coupling constant of the marginal perturbation.
When the perturbation is diagonal in the original currents of the theory,
there is a linear relation between the two, and non-linear otherwise. 
In the fermionic cases, both free and with Calogero-Sutherland interactions, the marginal perturbation
corresponds to backward scattering processes.
\end{abstract}
\vskip 0.5 cm
\end{titlepage}
\pagenumbering{arabic}
\section{Introduction} 

The Bogoliubov transformation is a powerful tool to find normal modes and
diagonalize interacting hamiltonians in many-body physics \cite{mattuck}. It plays a 
fundamental role in the development of the BCS theory of superconductivity
establishing that the main normal modes of the diagonal hamiltonian 
are quasiparticles that represent physically Cooper pairs. In other areas of many body physics it
has played a fundamental role for treating strong interactions as well, like in the 
realm of nuclear physics. 

In this paper, Bogoliubov transformations in systems 
of 1D spinless fermions are considered 
in the geometrical context of Hilbert space deformations. 
We consider systems in $(1+1)$ dimensions that posses a global $U(1)_L \times U(1)_R $ 
symmetry in their free form. 
Here L and R stand for left and right 
holomorphic coordinates, which also represent the sense of motion 
of the excitations described by the corresponding quantum fields. 
In the fermionic cases considered, the Fermi sea is topologically a segment 
and these are the symmetries that express 
particle number conservation around each Fermi point (we wound not
consider here coupling to an external electromagnetic field, that would
give rise to a charge anomaly).
For the class of theories considered, these symmetries are realized by 
a set of two currents, $J_L$ and $J_R$, of conformal dimensions
$(h,{\overline h})$ $(1,0)$ and $(0,1)$, respectively, meaning the
their scaling dimension $\delta = h+ {\overline h}= 1$ in both cases.
We refer to this as the decoupled basis, in which L and R sectors do not interact
with each other
and the theory is invariant under the ${\bf Z}_2 $ parity symmetry given by the
exchange of the L and R sectors. 
Still within the same theory, we may consider another basis for expressing
the same symmetry, which is written as  $U(1)_S \times U(1)_{AS} $, where 
we define a set of coupled currents:
$J_S = J_L + J_R $ and $J_{AS} = (J_R - J_L)/2 $. This basis couples 
both sectors and breaks parity explicitly, since $J_S $ 
is symmetric and $J_{AS} $ anti symmetric under it. 
In this paper we study the effects of 
including marginal $(h,{\overline h})=(1,1)$ interactions \cite{borsa}, 
and find that, according to the sign of its coupling, 
the theory deforms towards one of the coupled sectors,
$U(1)_S $ for positive couplings and $U(1)_{AS} $. 
The deformation itself can be described in terms of a Bogoliubov transformation,
interpreted geometrically as a Lorentz boost in the 2D space of
currents. 

In section 2 we consider the case of the free boson $(c,{\overline c})=(1,1)$  conformal field theory (CFT)
\cite{bpz,appcft}, and discuss its symmetries
and Hilbert space properties. We then study the deformation induced by the 
marginal coupling. In section 3, we consider the case of free Weyl fermions $(c,{\overline c})=(1,1)$  CFT in a physical
situation that describes two opposite moving waves on a circumference, and then investigate
the effects of the marginal deformation, which yield a deformation of the charge spectra,
in agreement with the free boson CFT. In section 4 we consider the $\winf \times {\overline \winf}$
extended CFT of the Calogero-Sutherland model, which is still a $(c,{\overline c})=(1,1)$ CFT
that differs from the free theory by the inclusion of interactions. We discuss an interesting phenomenon
turning an attractive small piece of the hamiltonian into the dominating one at low energies,
a process that has some analogy with the BCS mechanism. Finaly, we give some brief conclusions
in section 5.
 
\section{Free boson CFT} 

We first consider the $(c,{\overline c})=(1,1)$ CFT of a free bosonic compactified field $\Phi$~\cite{bpz} 
(our conventions follow \cite{appcft}) with euclidean action
\beq
S\ =\ \frac{1}{2\pi} \int d^2 z~\partial \Phi
\,{\overline\partial}\Phi~~~,
\label{bosac}
\eeq
and a compactification radius $r$ defined by $\Phi\ \equiv\ \Phi
+ 2\pi r$.
The classical equations of motion for $\Phi$, $\partial {\overline\partial}\Phi =0$  
are solved by 
\beq
\Phi(z,{\overline z})\ =\ \varphi(z)\ +\ {\overline\varphi}({\overline
z})~~~,
\label{clsol}
\eeq
where $\varphi(z)$ and ${\overline\varphi}({\overline z})$ are the holomorphic or
chiral and the anti-holomorphic or antichiral components of the boson field $\Phi$, respectively.
The solutions may be given as a Laurent series expansion of the form: 
\beq
\Phi(z,{\overline z})\ =\ \varphi_0\ -i\alpha_{0} \ln (z)\ -i{\overline \alpha}_{0} \ln ({\overline z})\ -\
i\sum_{n \neq 0} \frac{\alpha_n}{n}\ z^{n}\ +\ i\sum_{n \neq 0} \frac{{\overline \alpha}_n}{n}\ {\overline z}^{n}\ .
\label{phiexp}
\eeq
In the quantum theory, the coefficients $\varphi_0$, ${\alpha_n}{n}$ and ${\overline \alpha}_n$ become quantum 
operators. The details of the quantization procedure may be found in, {\it e.g.}, \cite{stone}.
The non-zero bosonic operators satisfy the commutation relation
\barr
\left[\ \alpha_{\ell},\alpha_{m}\ \right] &=& \ell\ \delta_{\ell+m,0}\ ,\nl
\left[\ {{\overline \alpha}}_{\ell},{\overline \alpha}_{m}\ \right] &=& \ell\ \delta_{\ell+m,0}\ ,
\label{oscal}
\earr
known as the oscillator algebra since $a_m = \alpha_m /\sqrt{m}$ and $a^{\dagger}_m = \alpha_{-m} /\sqrt{m}$ 
for $m> 0 $ are standard ladder harmonic oscillator operators, and analogously for the ${\overline \alpha}$ set. 
Well defined coordinates under maps $z \to {\rm e}^{2\pi i n_1}\ z $ and 
${\overline z} \to {\rm e}^{2\pi i n_2}\ {\overline z}$ imply that $n_1$ and $n_2$ should be integers
and that $\Phi(z,{\overline z}) \to  \Phi(z,{\overline z})+2\pi r (n_1 +n_2 )$. 
It follows that the spectrum of $-i\alpha_{0}$ and $i{\overline \alpha}_{0}$ should be given 
by $r n_1$ and $r n_2$, respectively. 
So far, the quantum field (\ref{phiexp}) does not split in chirality as the classical solution (\ref{clsol})
does. To achive this, a couple of steps are usually performed: {\it i)} define new zero-mode fields
by the symmetric and anti-symmetric combination of them, as in center of mass and relative coordinates
of two particles, and {\it ii)} introduce a new conjugate field so as to complement the coordinate $\varphi_0 $
\cite{stone}.
\beq
P\ =\  \alpha_{0}\ + {\overline \alpha}_{0}\ , \qquad
N\ =\  \frac{1}{2} \left( \alpha_{0} -  {\overline \alpha}_{0} \right)\ .
\label{pandn}
\eeq
The first operator is called the momentum as is conjugate to $\varphi_0 $, with spectrum given by
$m/(2r)$, where $m$ is an integer. The second is called the winding number and its spectrum is
given by $nr$, with integer $r$ (more details are to be found in \cite{appcft} as well). 
The new field to be introduced is called $\theta$, and conjugate to $N$. 
The chirally independent fields become:
\barr
\varphi (z) &=& \frac{1}{2}\ \varphi_0\ +\ \frac{1}{r}\ \theta \ -\
i\alpha_{0} \ln (z)\ +\ i\sum_{n \neq 0} \frac{\alpha_n}{n}\ z^{n}\ ,\nl
{\overline \varphi}({\overline z}) &=& \frac{1}{2}\ \varphi_0\ -\ \frac{1}{r}\ \theta\ -\ 
i{\overline \alpha}_{0} \ln ({\overline z})\ -\
i\sum_{n \neq 0} \frac{{\overline \alpha}_n}{n}\ {\overline z}^{n}\  .
\label{auansp}
\earr
The spectra of charges of $\alpha_{0}$ and ${\overline \alpha}_{0}$ are given 
by $(m/(2r) +nr)$ and $(m/(2r) -nr)$, respectively. 

Notice that under the spatial ${\bf Z}_2$ parity symmetry (holomorphic-antiholomorphic exchange)
the action (\ref{bosac}) is invariant, as well as the 
the classical solutions (\ref{clsol}). However, it appears to be broken by the quantization
conditions which include the odd operator $N$ in (\ref{pandn}). 
However, one can define another free boson field, the so-called dual field denoted 
as ${\tilde \Phi}$ or $\Theta$ obeying the same action (\ref{bosac}), but such that 
the currents are the dual of the previous ones defined for $\Phi$. 
\beq
\Theta(z,{\overline z})\ =\ \frac{1}{2}\ \left[ \varphi(z)\ -\ {\overline\varphi}({\overline z}) \right]~~~,
\label{thetaf}
\eeq
where the chiral components are the same as those for the $\Phi$ field, (\ref{auansp}). 
The compactification radius for $\Theta$ is $2/r$. The T(for target)-duality  is a 
${\bf Z}_2$ symmetry of the Hilbert space of the free boson CFT that refers to
the exchange between $\Phi$ and $\Theta$, the former asociated to electric (symmetric) excitations 
and the latter to magnetic (antisymmetric) ones. 
Since the symmetry $r \leftrightarrow 1/(2r)$ is 
manifest in the spectrum of charges of (\ref{bosac}),the boson field $\Phi$ 
may be identified with $\Theta$ such that both symmetric and anti-symmetric combinations
of currents are independent fields. This symmetry is then also known as 
`electric-magnetic' duality: $r \leftrightarrow 1/(2r)$, $\partial \Phi {\overline\partial}\Phi \leftrightarrow  
- \partial \Phi {\overline\partial}\Phi $ and $V_{m,n} \leftrightarrow V_{n,m}$. 
The partition function of the $(c,{\overline c})=(1,1)$ free boson theory is invariant under 
the `electric-magnetic' duality, because under it $Q \leftrightarrow Q$, ${\overline Q} 
\leftrightarrow -{\overline Q} $, but conformal dimensions, quadratic in these quantities, are
invariant and the latter is the data that enters in the determination of the partition function \cite{dijk,ginsp}.

The chiral and antichiral currents, of conformal dimensions $(1,0)$ 
and $(0,1)$, respectively, are given by:
$J(z) = \partial \varphi(z) $ and ${\overline J}(z) = {\overline \partial} {\overline \varphi}({\overline z}) $.
The Laurent modes of these current on the $z$ plane are:
\beq
J_n\ =\ \oint\ \frac{dz}{2\pi i}\ z^{n}\ J(z)\ , \qquad 
{\overline J}_n\ =\ \oint\ \frac{d{\overline z}}{2\pi i}\ {\overline z}^{n}\ {\overline J}({\overline z})\  .
\label{fourj}
\eeq
Moreover, the stress-energy tensor for this system has only chiral components, as it
is a CFT, and given by the Sugawara construction \cite{bpz}:
\beq
T(z)\ =\ {\frac{1}{2\xi}}\ :\left( J (z)\right)^2 :\ , \qquad 
{\overline T}({\overline z})\ =\ {\frac{1}{2\xi}}\ :\left( {\overline J}({\overline z})\right)^2 :\ ,
\label{stress}
\eeq
with $\xi$ a real parameter paremeter. The Laurent modes
\beq
L_n\ =\ \oint\ \frac{dz}{2\pi i}\ z^{n+1}\ T(z)\ , \qquad 
{\overline L}_n\ =\ \oint\ \frac{d{\overline z}}{2\pi i}\ {\overline z}^{n+1}\ {\overline T}({\overline z})\  .
\label{fourt}
\eeq
The modes of these operators satisfy the abelian current algebra:
\barr
\left[\ J_{\ell},J_{m}\ \right] & = &  c\ \xi \ell\ \delta_{\ell+m,0} ~~~,\nl
\left[\ L_{\ell}, J_m\ \right] & = & -m\ J_{\ell+m} ~~~,\nl
\left[\ L_\ell, L_m\ \right] & = & (\ell-m)L_{\ell+m} + 
\frac{c}{12}\ell(\ell^2-1) \delta_{\ell+m,0}~~~,
\label{kmalg1}
\earr
with $c=1$ and where $\xi$ is the level, a real free parameter, that is practice is a function 
of the compactification radius (more specifically, $\xi = 1/r^2 $).  
A comparison of (\ref{kmalg1}) with (\ref{oscal}) shows that $ J_n = \alpha_n$ for $\xi=1$.
The spectrum of these currents can be given
in terms of two numbers $m$ and $n$, with
$m$ integer and $n$ integer (for $m$ even) or half-integer (for $m$ odd).
In particular, the highest weight states denoted by
$|\,m,n\,\rangle$, are such that
\barr
J_0\,|\,m,n\,\rangle
&=& Q\, |\,m,n\,\rangle  ~~~,\nl
{\overline J}_0\,|\,m,n\,\rangle &=& {\overline Q}\,
|\,m,n\,\rangle~~~,
\label{specbo'}
\earr
where the chiral and antichiral charges $Q$ and ${\overline Q}$ are,
respectively
\beq
Q = \frac{m}{2r} + r\,n ~~~~,~~~~{\overline Q} = \frac{m}{2r}
-r\,n~~~.
\label{specbo}
\eeq
The terms of the form $m/2r$ are know as 'electric' charges, whereas those of type $nr$ are known as 'magnetic'. 
The quantity $p=m/r$ is usually called the `momentum' and $w= nr$, the `winding mode´. 
The states $|\,m,n\,\rangle$ are created by vertex operators
\beq
V_{n,m} (z, {\overline z}) \ =\ \exp\left[ iQ\ \varphi( z)\ +\ i{\overline Q}\ {\overline \varphi}
({\overline z}) \right]\ , 
\label{vertexo}
\eeq
by $|\,m,n\,\rangle \ =\ V_{n,m} (0,0) |\,0,0\,\rangle$, where $|\,0,0\,\rangle$ is the $SL(2,{\bf C})$
vacuum.
Note that the two chiral components of $\Phi$ are not totally
independent since there is an overall constraint of charge
conservation. Furthermore, for general values of $r$,
the charges $Q$ and ${\overline Q}$
are not necessarily integers, as opposed to the case $r = 1$, that describes free Weyl fermions.
The spectra of conformal dimensions are given by:
\barr
L_0\,|\,m,n\,\rangle
&=& h\, |\,m,n\,\rangle  ~~~,\nl
{\overline L}_0\,|\,m,n\,\rangle &=& {\overline h}\,
|\,m,n\,\rangle~~~.
\label{spechd}
\earr

\subsection{Marginal $J-{\overline J}$ perturbations}

\begin{figure}[t]
\begin{center}
\includegraphics[width=14cm]{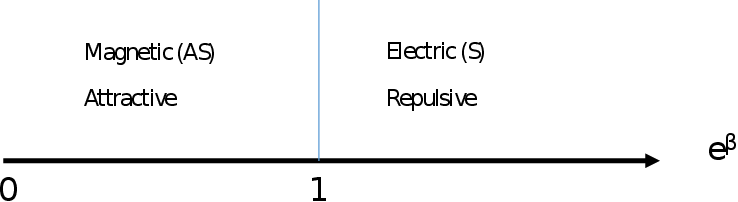}
\end{center}
\caption{Landscape of the regions explored by the free boson CFT as a 
function of the deformation parameter ${\rm e}^{\beta}$.
The properties characterizing each region are not exclusive 
but dominant and the diagram does not describe different
phases.}
\label{defo-fig}
\end{figure}

Now we consider perturbations to the free action (\ref{bosac}) that are given by marginal operators.
These deformations preserve conformal symmetry and the central
charges, and are generated by fields ${\cal O} (z, {\overline z})$
of conformal dimensions $(1,1)$. 
A prime example of $(1,1)$ operator is ${\cal O} (z, {\overline z})\ =\ \partial \Phi
\,{\overline\partial}\Phi~$. For a small value of the dimensionless
coupling constant $\delta g$, the free action (\ref{bosac}) is modified by the addition of the 
term $\delta S $ as $S \to S + \delta S$, with 
\beq
\delta S\ =\ -\delta g\ \int\ d^2z\ \partial \Phi
\,{\overline\partial}\Phi\ .
\label{deltas}
\eeq
This perturbation modifies the overall normalization
of the action $S$, which may be absorbed by a redefinition of the field 
$\Phi \to  {\hat \Phi} = \sqrt{1-\delta g}\ \Phi$ so that
the action returns to its original form in terms of the new field. 
This implies a change in the compactification radius of the field
${\hat \Phi}$ given by $\delta r^2 = -\delta g\ r^2$. 
This is a first order perturbative formulation. However, fields
may have corrections in their conformal weights by perturbative
corrections. 
The condition for the conformal weights of the $(1,1)$ operators to be free from perturbative 
corrections have been studied in \cite{kada} , and is automatically satisfied in this
simple example \cite{dijk,ginsp}.
Considering the perturbative equation for $r^2 $ as a differential equation,
we find the following 
expression as the solution with undeformed value $r_{0}=1$ :
\beq 
r\ =\ {\rm e}^{ -\beta}\ ,
\label{radg}
\eeq
in terms of the deformation parameter $\beta = g/2$. 
For $\beta > 0$ the radius diminishes and it growths for $\beta < 0$.
This modification only affects the zero mode of the boson field,
and therefore does not change the action. However, the spectrum is modified 
accordingly by factors dependent on $r$. 
The deformed spectra of charges (\ref{specbo}) are given by;
\beq
{\hat Q} = \frac{m}{2 }{\rm e}^{\beta} + n {\rm e}^{-\beta}~~~~,~~~~
{\hat {\overline Q}} = \frac{m}{2}{\rm e}^{\beta}-n{\rm e}^{-\beta}~~~.
\label{specbo1}
\eeq
The spectrum of conformal dimensions is given by 
\beq
{\hat h}\ =\ \frac{1}{2\xi}\ {\hat Q}^{2} = \frac{r^2}{2}\ \left( \frac{m}{2 r_{0} }{\rm e}^{-\beta} + 
n r_{0}{\rm e}^{\beta}\right)^{2}~~~~,~~~~{\hat {\overline h}}\ 
=\ \frac{1}{2\xi}\ {\hat{\overline Q}}^{2}\ =\ \frac{r^2}{2}\ \left( \frac{m}{2r_{0}}{\rm e}^{-\beta}
-nr_{0}{\rm e}^{\beta} \right)^{2}~~~.
\label{specbo2}
\eeq
It satifies that ${\hat h}- {\hat {\overline h}}= h - {\overline h} = r^{2} m n$. 
This means that the momentum
operator $L_{0} - {\overline L}_{0}$ is invariant under the flow of $r$, up to a scale
factor.
The relationship between the charge spectra before and after the deformation is given by 
an inverse {\it Lorentz boost} in $(1+1)$ dimensions, that is a $SO^{+}(1,1)$ transformation in the 
2D space of currents: 
\beq
\left[ \matrix{ {\hat Q}   \cr
{\overline {\hat Q}}  \cr} \right]\ =\
\left[ \matrix{ \cosh(\beta ) & \sinh(\beta )  \cr
\sinh(\beta ) & \cosh(\beta ) \cr} \right] 
\left[ \matrix{ Q   \cr
{\overline Q}  \cr} \right] \ .
\label{bogo22}
\eeq
This is a bosonic Bogoliubov tranformation.
Here the `space´ coordinate is the $L$ component, whereas the `time´ ones is given by the $R$
component, and the deformation parameter, termed as rapidity, is related to the relative dimensionless `velocity´
by $\tanh (\beta) = v$.
The old coordinate system moves with velocity $-v$ with respect to the new one.  
Had we considered the symmetric and anti-symmetric basis of charges, $Q_{S} = Q + {\overline Q}$ and
$Q_{AS} = (Q - {\overline Q})/2$, we would have obtained, instead, a 
diagonal form of the above  $SO^{+}(1,1)$ transformation:
\beq
\left[ \matrix{ {\hat Q}_{S}   \cr
{\overline {\hat Q}}_{AS}  \cr} \right]\ =\
\left[ \matrix{ {\rm e}^{\beta } & 0  \cr
0 & {\rm e}^{-\beta } \cr} \right] 
\left[ \matrix{ Q_{S}   \cr
{\overline Q}_{AS}  \cr} \right] \ .
\label{bogoonq}
\eeq
This is known as a squeeze mapping of parameter ${\rm e}^{\beta}$.
For $\beta > 0 $, charges are projected on the S sector (electric, or repulsive), with the opposite
behavior for $\beta < 0 $ (magnetic or attractive) (see Fig. \ref{defo-fig}).
Similarly, the 2D lattice of charges, is generated by 
the vectors $\vec{v_{s}}(r) = \left( \frac{1}{2r}, 0 \right)$ and $\vec{v_{as}}(r) = \left( 0, r \right)$
(the subindex $s$ stands for symmetric and $as$ for anti symmetric, as in (\ref{specbo})).
These
evolve with $\beta$ as: $\vec{v_{s}}(r) = {\rm e}^{\beta} \vec{v_{s}}(r_{0})$ and 
$\vec{v_{as}}(r) = {\rm e}^{-\beta} \vec{v_{as}}(r_{0})$. We note that the area of the unit cell is
constant and equal to $1/2$, and independent of the deformation. For large $\beta> 0$, the deformation
yields a projection onto the S sector, with the opposite results for $\beta < 0$.
\beq
\left[ \matrix{ 1/(2\hat r )   \cr
{\hat r}  \cr} \right]\ =\
\left[ \matrix{ {\rm e}^{\beta} & 0  \cr
0 & {\rm e}^{-\beta} \cr} \right] 
\left[ \matrix{ 1/(2r)   \cr
r  \cr} \right]
\label{boost1}
\eeq
Notice that chiral splitting is preserved in the deformed CFT:
\beq
{\hat \Phi} (z,{\overline z})\ = {\hat\varphi }(z) + {\hat {\overline\varphi}}({\overline
z})~~~,
\label{chspl}
\eeq
where ${\hat\varphi }(z)$ and ${\hat {\overline\varphi}} ({\overline z})$ are the
chiral
and antichiral components of the boson field ${\hat \Phi}$, respectively.

For $r=1$, charges becomes half integer, and for $m=0$, the charge spectrum is given by 
integer values $Q=n$ and ${\overline Q} =-n$, with conformal dimensions corresponding to these excitations 
given by $h =  n^{2}/2 $ and
$h = {\overline n}^{2}/2 $, respectively. For $n=\pm 1$ these are fermion fields, as the relation 
of the quantum statistics $\theta$ with conformal dimensions is given by $\theta / \pi = 2h$.
These excitations are $(1+1)$-dimensional relativistic (massless) chiral charged fermions
and are identified, therefore, as Weyl fermions. In terms of the vertex operators 
(\ref{vertexo}), these correspond to $V_{0,\pm 1}$. Explicitly, 
\beq
F(z)\,\equiv \ :\,\exp ( i\varphi(z)):
{}~~~,~~~
F^\dagger(z)\, \equiv\ :\,\exp (- i\varphi(z)):~~~.
\eeq
Notice that under the deformation (\ref{radg}), these excitations acquire charges ${\hat Q}= n {\rm e}^{-\beta}$
and conformal dimensions $h = n^{2}{\rm e}^{-4\beta}/2 $ and therefore are no longer 
fermions. 
In many applications, fermionic formulations of the Weyl CFT are widely used.

To summarize, the effect of the one-parameter marginal deformation (\ref{deltas}) on the free boson CFT
is to deform the spectra of charges and conformal dimensions (\ref{specbo1},\ref{specbo2}). The geometry
of it is given by a Bogoliubov transformation, interpreted as a Lorentz boost in the 2D space 
of L-R currents. The value of the deformation parameter is not fixed as there is not an interaction
scale.

\section{Free Weyl fermion CFT}

We now consider fermionic degrees of freedom and follow a line of reasoning similar 
to that of the previous section. 
In many physical applications, operators and fields are
naturally defined on a spatially compact space, like a circumference 
of radius $R$. The corresponding Minkowskian theory is
then defined on the cylinder formed by the spatial circumference times the real line
that represents the time coordinate. Therefore, the coordinates for this class
of systems are $(x,t)$. 
In terms of two sets of anticommuting Fock operators 
$\{ (a^{s})^\dagger_n, a^{s}_m \} = \delta_{n+m,0}$, $\{ a^{s}_n, a^{s}_m \} = 0$, 
$\{ (a^{s})^\dagger_n, (a^{s})^\dagger_m \} = 0$, with $s=L,R$,
it is natural to define the following two chiral Weyl fermions (see, {\it e.g.}, \cite{flsz}),
that describe the low energy degrees of freedom of a system of fermions around 
each Fermi point, $R$ and $L$:
\beq
F_{R}(\theta) = \frac{1}{\sqrt{R}}\sum_{r=-\infty}^\infty
a^{R}_r~{\rm e}^{i\left(r-\frac{1}{2}\right)\theta}~~~,
\label{f+}
\eeq
and
\beq
F_{L}(\theta) = \frac{1}{\sqrt{R}}\sum_{r=-\infty}^\infty
a^{L}_r~{\rm e}^{i\left(r-\frac{1}{2}\right)\theta}~~~.
\label{f-}
\eeq
Since the index $r$ is integer, these fields
satisfy antiperiodic boundary conditions on a
circle of radius $R$, parametrized by the angle 
$\theta =x/R$. Introducing a dimensionless velocity parameter, their free hamiltonian $H_0$ can be written as
\beq
H_0\, =\ \frac{v}{R}\int_0^{2\pi}d\theta~
:\,\left[F_{R}^\dagger\,(-i\partial_\theta)F_{R}\ +\
F_{L}^\dagger\,(-i\partial_\theta)F_{L}\right]\,:~~~.
\label{dirac}
\eeq
This is the Dirac hamiltonian for a free
relativistic (massless) fermion in $(1+1)$-dimensions given by
\beq
\Phi=\left(\matrix{
 F_{R} \cr
 F_{L} \cr}\right)~~~.
\label{diraweyl}
\eeq
As it is well-known, in $(1+1)$-dimensions a Dirac fermion is
equivalent to a scalar boson through the abelian bosonization
procedure.
Thus, the free hamiltonian $H_0$ can be given an equivalent
description using bosons instead of fermions.
Actually, we shall see that this is possible
even for the interaction hamiltonian $H_I$.

On the other hand, in the CFT literature operators and
fields are conventionally defined on the complex plane.
Therefore, it is convenient to map the physical operators from
the cylinder to the plane, where one may use
the mathematical results.
There is a well-known conformal mapping between the cylinder
$(u=\tau -iR\theta)$, and the conformal plane $(z)$,
namely
\beq
z=\exp\left({u\over R}\right)=
\exp\left({\tau\over R} -i\theta \right) ~~~,
\label{cft}
\eeq
where $\tau$ denotes the euclidean time.
Under this map, the Weyl fermion (\ref{f+}), which is
a primary field of weight $h=1/2$, takes the form (at $\tau=0$)
\barr
F(z) &=& \left({du\over dz}\right)^{1/2}  F_{+}(\theta) =
\sum_{r=-\infty}^{\infty} {\rm e}^{i\, r\,\theta}\ a_r ~~~, \nl
F^\dagger(z) &=& \left({du\over dz}\right)^{1/2}
F^\dagger_{+}(\theta)
=\sum_{r=-\infty}^{\infty} {\rm e}^{-i\,(r-1)\,\theta}\
a^\dagger_r ~~~.
\label{weypl}
\earr
Notice the well-known fact that due to the map (\ref{cft}), the
definition of $F^\dagger(z)$ differs from the naive expression.
We have suppressed the supraindex $(R)$ in (\ref{weypl}) for simplicity,
{\it i.e.}, $a_r \equiv a^{R}_r$.
The expression for the antichiral fields can be simply obtained from
(\ref{weypl}) by replacing $a_r$ with $a^{L}_r$, and $\theta$ with
$-\theta$, {\it i.e.} $z$ with ${\overline z}$.

The Fock space expressions of the generators for
the chiral current and Virasoro modes are:
\barr
J_n &=& \sum_{r=-\infty}^{\infty}: a^\dagger_{r-n}\, a_r :
{}~~~,\nl
L_n &=& \sum_{r=-\infty}^{\infty} \left(\,r-{n+1\over 2}\,\right)
:  a^\dagger_{r-n}\, a_r :~~~,\
\label{weylfo}
\earr
On the complex plane, the corresponding currents and the stress energy tensor, determined by the 
Sugawara construction, can then be written as follows
\barr
J(z) &=& :F^\dagger\,F: ~~~,\nl
T(z) &=& {1\over 2}\, :\, \partial \left(F^\dagger\,F\right):
-:F^\dagger\,\partial F: ~~~,
\label{wcurp}
\earr
where colons denote the normal ordering with respect to the ground
state.

\section{Weyl fermion CFT with marginal $J-{\overline J}$ perturbations}

\subsection{The perturbed relativistic rotor}

We would like to discuss the effects of perturbing the free Weyl fermion CFT by a marginal operator
of the form $J(z) {\overline J} ({\overline z})$. 
First, we consider a CFT that is analogous to the deformed free boson CFT of section 2. 
We notice, however, that this
interaction may not have the correct transformation property under the $L \leftrightarrow R$
exchange, as we consider a global hamiltonian factor, including the perturbation, 
proportional to the Fermi velocity.  
Thus, let us consider a system with hamiltonian given by:
\beq
H_{0}\ =\ \frac{v}{R}\ \left( L_0 \ -\ {\overline L}_0\right) \ ,
\label{luttih0}
\eeq
where $L_0 $ and ${\overline L}_0$ are Virasoro operators. 
It represents the relativistic (in the sense of linearity of the dispersion relation) 
kinetic energy of a system of two counter-propagating linear waves on a space-time cylinder 
made out of a spatial circumference of radius $R$ and the time coordinate. The
velocity of the waves has the same magnitude, such that the hamiltonian
(\ref{luttih0}) preserves total angular momentum in the cylnder, which in turn is the momentum
$(L_0 \ -\ {\overline L}_0)$ on the complex plane. 
Notice that this hamiltonian is parity invariant, provided $v \to -v$ under 
a L-R exchange. 
Moreover, (\ref{luttih0}) is not bounded from below, and therefore its spectrum 
cannot be interpreted as that of a {\it bona fide} physical system. However,
it is precisely what is needed to make a fermionic analogy of the CFT of section 2.
Moreover, this system may be thought
as a Luttinger model, albeit under the conditions just discussed. 
In the CFT language, it is given by the direct
product of two free chiral boson fields, such that the combined
theory has central charges $(c, {\overline c})=(1,1)$. 
Using the Sugawara construction, $T(z) = :J^{2} (z):/2$ and  
${\overline T}({\overline z}) = :{\overline J}^{2} ({\overline z}):/2$, 
the hamiltonian density may be written as:
\beq
{\cal H}_{0}(z,{\overline z}) \ =\ \frac{v}{2R}\ \left( :J^{2} (z): \ -\  
:{\overline J}^{2} ({\overline z}):\right)\ .
\label{lutth00}
\eeq
The connection with the bosonic CFT of the previous section is obtained
identifying $J(z) = \partial \varphi $ and 
${\overline J}({\overline z}) = {\overline \partial} {\overline \varphi} $. 
In order to study the effects of a marginal perturbation of (\ref{lutth0}) inducing
an interaction between the R and L Fermi points, is is better to consider the coupled
currents basis, so as to have consider the effects under parity transformations.
We define a basis of symmetric and anti symmetric currents as:
\beq
\left[ \matrix{ J_S   \cr
J_{AS}  \cr} \right]\ =\ 
\left[ \matrix{ 1 & 1  \cr
1/2 & -1/2 \cr} \right] 
\left[ \matrix{ J   \cr
{\overline J} \cr} \right]
\label{jjsas}
\eeq
This basis is closely related to that of the free boson field and
its dual in terms of its chiral current components. 
Since (\ref{lutth00}) has a hyperbolic structure in terms of the currents,
it is natural to consider transformations in the $2D$ space of currents 
that preserve it. These are given by $SO^{+}(1,1)$ transformations,
which may be written generically as: 
\beq
\left[ \matrix{ {\hat J}   \cr
{\hat {\overline J}}  \cr} \right]\ =\
\left[ \matrix{ \cosh(\beta ) & \sinh(\beta )  \cr
\sinh(\beta ) & \cosh(\beta ) \cr} \right] 
\left[ \matrix{ J   \cr
{\overline J}  \cr} \right]
\label{bogo0}
\eeq
where $\beta$ is the rapidity (the angle of the imaginary rotation).
This transformation preserves the hyperbolae formed by these pairs of currents:
\beq
{\hat J}^{2} \ -\ {\overline {\hat J}}^{2}\ =\ J^{2} \ -\ {\overline J}^{2}\ .
\label{hyper}
\eeq
The way of selecting $\beta$ depends on the physical conditions desired. 
For the moment, we will consider it as a free parameter.

Bogoliubov transformations are considered an isomorphisms between bosonic 
or fermionic Fock operators, therefore preserving the original commutation
or anticommutation relations. It is frequently employed in many-body physics to diagonalize 
hamiltonians with some types of interactions \cite{mattuck}. 
In terms of the bosonic current modes defined in (\ref{fourj}), 
a generic Bogoliubov transformation reads:
\beq
\left[ \matrix{ {\hat J}_{\ell}   \cr
{\hat {\overline J}}_{-\ell}  \cr} \right]\ =\
\left[ \matrix{ \cosh(\beta ) & \sinh(\beta )  \cr
\sinh(\beta ) & \cosh(\beta ) \cr} \right] 
\left[ \matrix{ J_{\ell}   \cr
{\overline J}_{-\ell}  \cr} \right]\ .
\label{bogomode}
\eeq
These current modes may be integrated to define current densities 
by the relations:
\beq
J(z)\ =\ \sum_n\ J_n\ z^{-n-1}\ ,\qquad {\overline J}(z)\ =\ \sum_n\ {\overline J}_n\ z^{-n-1}\ .
\label{defcubo}
\eeq
The second expression is understood as a formal one, since the actual dependence of ${\overline J}$
is on ${\overline z}$ (see also \cite{dege} for a similar treatment).
Therefore, (\ref{bogo0}) may be considered as a generic Bogoliubov transformation for the
class of systems we are considering. 
Notice that in terms of the coupled currents, (\ref{bogo0}) reads as:
\beq
\left[ \matrix{ {\hat J_{S}}   \cr
{\hat {\overline J_{AS}}}  \cr} \right]\ =\ 
\left[ \matrix{ 
{\rm e}^{\beta}
& 0  \cr
0 & {\rm e}^{-\beta} \cr} \right] 
\left[ \matrix{ J_{S}   \cr
J_{AS} \cr} \right]
\label{bogo04}
\eeq
Therefore, as $|\beta| \to +\infty$, (\ref{bogo04}) acts as
a projector on  $J_{S}$ for $\beta > 0$,  and on
$J_{AS}$ for $\beta < 0$.
In the coupled basis, the free hamiltonian density (\ref{lutth00}) reads:
\beq
{\cal H}_{0}(z,{\overline z}) \ =\ \frac{v}{R}\ J_{S}\ J_{AS}\ .
\label{lutth1}
\eeq
Now we perturbe this density by adding an interaction term of the form:
\beq
\delta {\cal H}_I\ =\  \frac{\delta g\ v}{R}\ J_{S}\ J_{AS}\ ,
\label{perth}
\eeq
where $\delta g$ is a dimensionless coupling constant. 
The marginal deformation considered is of the class of the interaction terms induced by backward
scattering in Tomonaga-Luttinger modes, for example \cite{voit,lueme,solyo}.
As in the case of the free boson action, the combined hamiltonian has the
same form as the original one, except for an overall scale change.
We define new currents ${\hat J}_{S} = (1+\delta g) J_{S}$
and ${\hat J}_{AS} = J_{AS}$. Notice that this amounts to deform the total particle number current 
and not the dual one, so as to preserve the L-R parity symmetry as in the case of the free boson.
The total hamiltonian density ${\cal H} ={\cal H}_{0} + \delta {\cal H}  $ in terms of 
the new currents becomes:
\beq
{\cal H}(z,{\overline z}) \ =\ \frac{v}{2R}\ \left( :{\hat J}^{2} (z): \ -\ 
:{\overline {\hat J}}^{2} ({\overline z}):\right) \ ,
\label{lutth0}
\eeq
that it, it returns to the form of ${\cal H}_{0}$. In terms of the original chiral currents, we have:
\beq
\left[ \matrix{ {\hat J}   \cr
{\hat {\overline J}}  \cr} \right]\ =\
\left[ \matrix{ (1+ \delta g /2) & \delta g /2  \cr
\delta g /2 & (1+ \delta g /2) \cr} \right] 
\left[ \matrix{ J   \cr
{\overline J}  \cr} \right]
\label{bogo3}
\eeq
which is of the form of Bogoliubov transformation (\ref{bogo0})
for small deformation parameter $\beta = g/2$. 
The finite deformation is given by the parameter ${\rm e}^{\beta}$.
In terms of it, the 
transformation matrix in (\ref{bogo3}) becomes that of (\ref{bogo0}). 
The hamiltonian $H=H_{0}+H_{I}$, with $H_I = g H_0$ becomes 
${\hat H} = {\rm e}^{\beta} H_0 $, with $J_{S} \dot J_{AS}= {\hat J}_{S} \dot {\hat J}_{AS}$ and
${\hat J}_{S} = {\rm e}^{\beta} J_{S}$, ${\hat J}_{AS} = {\rm e}^{-\beta} J_{AS}$. 
We are therefore led to conclude that $v$ transforms as well as ${\hat v} = {\rm e}^{\beta} v$.
In terms of the transformed variables, the hamiltonian is completely isomorphic
to the unperturbed one: 
\beq
{\hat H}\ =\ \frac{\hat v}{R}\ \left( {\hat L}_0 \ -\ {\hat {\overline L}}_0\right) \ .
\label{luttfree}
\eeq
The action of the Bogoliubov transformation on the marginally perturbed hamiltonian may be thought of 
as a diagonalization as well, in the sense of providing a realization of a new
hamiltonian isomorphic to the free one. Marginal interactions are, consequently,
not {\it bona fide} ones, but maps between isomorphic free Hilbert spaces. 
It is clear, however, than the unperturbed and perturbed hamiltonian are both diagonal 
in their respective bases.

We therefore obtain a result that is completely analogous to that of the free bosonic case:
the spectra of charges and conformal dimensions are given by (\ref{specbo1}) and (\ref{specbo2}),
respectively. This result is clearly expected given the bosonization rules,
but this discussion helps in extending it to other systems, as it will be clear in a moment.
The effect of the marginal deformation is to map the original Hilbert space 
of the theory onto a deformed one, isomorphic to the original. 
The quantitative differences only apply to the values of their 
spectra \cite{cardy}. 
Therefore, the most important action of this deformation is on the Hilbert space.
The process is compatible with the L-R symmetry. 
This result obtained in hamiltonian language, closer to condensed 
matter formulations, is equivalent to the previously discussed one for the free boson field. 

\subsection{The Bogoliubov transformation in Hilbert space}

Following the previous analysis, we now focus on the effect of the Bogoliubov transformation
on the free Hilbert space.
There should exist a similarity transformation
$T(\beta)$ that realizes the transformation as follows \cite{halb}:
\beq
{\hat J}_{\ell}\ =\  T(\beta)\ J_{\ell}\ T^{-1}(\beta)
\label{bogohil}
\eeq
for all $\ell$. 
Given the transformation in the 2D space of currents: 
\beq
{\hat J}_{\ell}\ =\ \cosh( \beta )\ J_{\ell}\  +\  \sinh (\beta )\ {\overline J}_{-\ell}\ 
\label{bogo1}
\eeq
On general  grounds, keeping the symmetry of momentum conservation, an ansatz for $T(\beta)$
is:
\beq
 T(\beta)\ =\ \exp\left( -\beta \sum_n\ C_n\ J_{n}{\overline J}_{-n}\ \right )\ ,
\label{anst}
\eeq
where $C_n $ are coefficients to be determined. 
Applying the Baker-Campbell-Hausdorff formula to expand (\ref{bogohil}) in powers of $\beta$,
and matching term to term the corresponding one of (\ref{bogo1}), we find $C_n = -1/n$ for $n \neq 0$. Therefore,
\beq
T(\beta)\ =\ \exp\left( \beta \sum_{n \neq 0}\ \frac{1}{n}\ J_{n}{\overline J}_{-n}\ \right)\ .
\label{anst2}
\eeq
We may use this form of $T$ to express the transformation of the free theory ground state to the
new highly correlated one:
\beq
|\ {\hat 0}\ ,\ {\hat 0}\ \rangle \ =\  \exp\left( \beta \sum_{n = 1}^{\infty}\ \frac{1}{n}\ J_{n}{\overline J}_{-n}\ \right)\  
|\ 0\ ,\ 0\ \rangle\ .
\label{anst3}
\eeq
More detailed and fundamental discussions may be found in \cite{dege,alga,forrog}.

\subsection{The Tomonaga-Luttinger model with backward scattering interactions}

Within the same CFT of free Weyl fermions, we now consider the system with hamiltonian 
\beq
H_{0}\ =\ \frac{v}{R}\ \left( L_0 \ +\ {\overline L}_0\right) \ ,
\label{lutt10}
\eeq
where $L_0 $ and ${\overline L}_0$ are Virasoro operators. It differs from (\ref{luttih0})
in the sign of the ${\overline L}_0$, making the system bounded from below, and therefore
physically acceptable, but with free hamiltonian non longer invariant under the boost (\ref{bogo0}).
Actually, this hamiltonian is invariant under a $SO(2)$ rotation, implying an imaginary
rapidity in (\ref{bogo0}). Nevertheless, this is the system that is mostly associated 
to condensed matter applications, the Tomonaga-Luttinger model \cite{voit}. 
Under a parity L-R transformation, $v$ is invariant, and we may perturb the system 
with a simple marginal $J-{\overline J}$ term. We therefore add to the free hamiltonian
density
\beq
{\cal H}_{0} \ =\ \frac{v}{2R}\ \left( J^{2} \ +\  
{\overline J}^{2} \right)\ ,
\label{lutth100}
\eeq
the interaction term:
\beq
{\cal H}_{I}\ =\ \frac{v}{2R}\ \left( g\ J{\overline J}  \right) \ .
\label{lutt101}
\eeq
The combined density ${\cal H}={\cal H}_{0}+{\cal H}_{I}$ is customarily diagonalized
by a Bogoliuobov trasformation (\ref{bogo0}). 
One obtains
\beq
{\cal H} \ =\ \frac{v}{2R}\ \left[ \left[ \cosh(2\beta) -(g/2)\sinh(2\beta) \right]
\left( {\hat J}^{2}  + {\hat {\overline J}}^{2} \right)+  
\left[ g\cosh(2\beta) -2\sinh(2\beta) \right] {\hat J}{\hat {\overline J}}
\right]\ .
\label{lutth101}
\eeq
This hamiltonian density is diagonal if $\tanh(2\beta) = g/2$. 
This means that ${\cal H}$ may only be diagonalized by a Bogoliubov transformation
provided $|g| \leq 2$, and that for small $g$, $2\beta \simeq g/2$. 
With this determination, we obtain
\beq
{\cal H} \ =\ \frac{v}{2R}\ \frac{ 
\left( {\hat J}^{2} \ +\ {\hat {\overline J}}^{2} \right)}{\cosh(2\beta)}\ =\
\frac{v}{2R} \sqrt{1-\frac{g^{2}}{4}} \left( {\hat J}^{2} \ +\ {\hat {\overline J}}^{2} \right)\ .
\label{lutth102}
\eeq
From the point of view of the previous discussions, we may still view the Bogoliubiov
transformation (\ref{bogo0}) as a Lorentz boost. The difference is that here the boost
rapidity $\beta$ is not proportional to $g$. Note that we may determine $\beta$ 
by a procedure that does invoke the diagonalization procedure of (\ref{lutth101}): 
one rewrites the matrix of the Lorentz boost of rapidity $2\beta$ (\ref{bogo22}) in the form:
\beq
\left[ \matrix{ \cosh(2\beta ) & \sinh(2\beta )  \cr
\sinh(2\beta ) & \cosh(2\beta ) \cr} \right]\ =\
\cosh(2\beta) \left[ \matrix{ 1 & \tanh(2\beta )  \cr
\tanh(2\beta ) & 1 \cr} \right]\ . 
\label{bogo32}
\eeq
Subsequently, one rewrites the hamiltonian density ${\cal H}$ in the 2D space of L-R currents, as 
a quadratic form in terms of matrices:
\beq
{\cal H}\ =\  \frac{v}{2R}\ 
\left[ \matrix{ J &  {\overline J}} \right]\   
\left[ \matrix{ 1 & g/2  \cr
g/2 & 1 \cr} \right]\ 
\left[ \matrix{ J   \cr
{\overline J}  \cr} \right]\ .
\label{bogo33}
\eeq
Comparing the dimensionless hamiltonian density matrix in (\ref{bogo33}) with the 
corresponding normalized boost matrix of (\ref{bogo32}), one finds that the off-diagonal
element of the first fixes the rapidity in the second by equating $g/2 = \tanh(2\beta)$.
Incidentally, note that the eigenvalues of the dimensionless hamiltonian matrix in (\ref{bogo33})
are $1\pm g/2$, with eigenvectors $[1 -1]$ and $[1 1]$, respectively. For the repulsive regime $g >0 $,
the lowest energy state is, therefore, the S combination, in agreement with the general landscape of
Fig. \ref{defo-fig}.

In conclusion, Bogoliubov transformations select a rapidity in terms of the coupling constant
of the marginal interaction. For free theories, the relationship is linear, whereas for 
interacting cases, it is not. 

\section{Calogero-Sutherland extended CFT}

We now consider the EFT of the Calogero-Sutherland model, as it is and extension of the
free fermionic theory we just discussed. In the thermodynamic limit, it describes
two sets of opposite moving, non-linear waves of the Benjamin-Ono type \cite{aw,abw,bz}. 
For completeness, we recall 
the basic features of the Sutherland model, given by
a system of $N$ non-relativistic 
$(1+1)$-dimensional spinless fermions on a circumference of length $L$, with 
Hamiltonian \cite{sut} (in units where $\hbar =1$ and $2m=1$, with $m$ being 
the mass of the particles)
\beq
h_{CS}=\sum_{j=1}^N\ \left( \frac{1}{i} \frac{\partial}{\partial x_j}\
\right)^2\ +\ g\ \frac{\pi^2}{L^2}\ \sum_{i<j}\ \frac{1}{\sin^2
(\pi(x_i-x_j)/L) }\ ,
\label{ham}
\eeq
where $x_i$ ($i=1,\dots,N$) is the coordinate of the $i$-th particle
along the circumference, and $g$, 
the dimensionless coupling constant. Ground state stability imposes $g \geq -1/2$,
with both attractive ($-1/2 \leq g < 0$) and repulsive ($0 < g$) regimes. A standard 
reparametrization of the coupling constant is given by $g=2 \xi ( \xi -1)$, 
so that $\xi \geq 0$, where $0 \leq \xi < 1$ is the attractive region and $1  < \xi $ 
the repulsive one. 

The EFT of this model has been studied by taking the thermodynamic limit of the system 
with hamiltonian (\ref{ham}). Our description is based on the
effective field theory \cite{polch} obtained in \cite{clz,flsz,cfslz}. Alternative descriptions 
of the EFT may be found in \cite{kaya,khve,amos,poly}.
It is obtained by rewriting the system dynamics in terms of fields
describing the low energy fluctuations of the 1D Fermi R and L points.
The suitable fields are obtained from non-relativistic fermionic free fields, 
and are two sets of relativistic Weyl fermion fields around each of the Fermi points,
as it has been discussed for the free fermion case. 
Actually, the Hilbert space of the EFT is isomorphic to that of the $(c,{\overline c})=(1,1)$ CFTs
described in the previous sections. The only difference with the Tomonaga-Luttinger system relies
in the non-linear effective hamiltonian. The entire EFT is formulated in terms of 
fields that that obey the $\winf$ algebra \cite{shen,kac1}.
The general form of the $\winf$ algebra satisfied by the operators $V^j_m$, with $j=0,1,2,\dots$ and $m$ integer
is:
\beq
\left[\ V^i_\ell, V^j_m\ \right] = (j\ell-im) V^{i+j-1}_{\ell+m}
+q(i,j,\ell,m)V^{i+j-3}_{\ell+m}
+\cdots +\delta^{ij}\delta_{\ell+m,0}\ c\ d(i,\ell) \ ,
\label{walg}
\eeq
where the structure constants $q(i,j,\ell,m)$ and $d(i,\ell)$ 
are polynomial in their arguments, $c$ is the central charge, 
and the dots denote a finite number of terms involving the operators 
$V^{i+j-2k}_{\ell+m}\ $.
The $c=1$ $\winf$ algebra can be realized by either fermionic or bosonic 
operators. In order to avoid confusion, we denote the first 
set of generators as $V^0_{\ell}$ and $W^0_{\ell}$. 
Both sets fulfill the same algebraic relations, except for the
zero mode: there is a  
different standard normalization of the Heisenberg algebra for fermionic and
bosonic realizations of the $\winf$ algebra, namely:
$\left[ V^0_\ell,V^0_m \right]  =  \ell \delta_{\ell+m,0}$ and 
$\left[ W^0_\ell,W^0_m \right]  =  \xi \ell\ \delta_{\ell+m,0}$.
For further details, please see \cite{flsz,cfslz}.
For our current purposes, the following identifications with the free fermionic
operators of the previous sections will be enough: 
$V^0_{\ell} = J_{\ell}$ and $V^1_{\ell} = L_{\ell}$ for the R sector with analogous 
expressions for the L one. 

The effective hamiltonian may be given as power series in $1/N$ in the $\winf$ basis 
according to \cite{flsz,cfslz}:
\beq
H_{CS}\ =\ E_{CS}\ \sum_{k=0}^\infty
\left(\frac{1}{N}\right)^k\,{\cal H}_{(k)}~~~, 
\label{series}
\eeq
where
\beq
H_{(0)} =  \frac{1}{4} (1+g) \left(\v^0_0 +
\vb^0_0\right)~~~,
\label{hcv0}
\eeq
\beq
H_{(1)} = \left(1+\frac{g}{2}\right) \left(\v^1_0 +
\vb^1_0\right)+\frac{g}{2} \sum_{\ell=-\infty}^{\infty}
\v^0_{\ell}\, \vb^0_{\ell}~~~,
\label{hcv1}
\eeq
\barr
H_{(2)} &=&
\left(1+\frac{g}{4}\right) \left(\v^2_0 +\vb^2_0\right)
-\frac{1}{12} (1+g) \left(\v^0_0+\vb^0_0\right) \nl
&&-\ \frac{g}{4}\sum_{\ell=-\infty}^{\infty} |\ell |
\left( \v^0_{\ell}\,\v^0_{-\ell}+
\vb^0_{-\ell} \,\vb^0_{\ell}+
2\,\v^0_{\ell} \,\vb^0_{\ell}\right)\nl
&&+\ \frac{g}{2}
\sum_{\ell=-\infty}^{\infty} \left( \v^1_{\ell}\,\vb^0_{\ell}+
\v^0_{\ell} \,\vb^1_{\ell} \right) ~~~.
\label{hcv2}
\earr
Here $E_{CS} = v \xi N/R$ is the global energy scale (on the cylinder), where 
$v = 2\pi n_0 $ is the Fermi velocity. This expressions have been derived in terms
of a realization of the $\winf$ algebra is terms of free fermions.
For the sake of completeness, we review here the behavior of the different terms of 
(\ref{series}). 

{\it i)} Order $1/N$: the quadratic form (upon use of the Sugawara formula) in the RHS of (\ref{hcv1}) can
be diagonalized  by the following Bogoliubov transformation
\beq
\left[ \matrix{ \w^0_{\ell}   \cr
\wb^0_{-\ell}  \cr} \right]\ =\
\left[ \matrix{ \cosh(\beta ) & \sinh(\beta )  \cr
\sinh(\beta ) & \cosh(\beta ) \cr} \right] 
\left[ \matrix{ \v^0_{\ell}   \cr
\vb^0_{-\ell}  \cr} \right]\ .
\label{bogocs}
\eeq
for all $\ell$, with $\tanh 2\beta =g/(2+g)$ (
the details of the calculations are given in the Appendix). 
This assumes that $g$ is small, and yields the solution ${\rm e}^{2\beta} = \sqrt{1+g}$ \cite{flsz,cfslz}. 
One may also consider the expression that includes larger values of $g$, as given in \cite{bz,boze3}: 
$\tanh 2\beta =(\xi -1)/(\xi + 1)$, with solution ${\rm e}^{2\beta} = \xi$.

{\it i)} Order $1/N^{2}$: we start by considering the terms in the first and third lines 
of (\ref{hcv2}) (the ones that do not involve momentum dependent interactions).
It is shown in the appendix that these terms are diagonalized again by
the Bogoliubov transformation (\ref{bogocs}) with $\tanh 2\beta =g/(2+g)$.
On the other hand, of the three terms in the second line of (\ref{hcv2}) 
only the chiral ones are diagonalized by (\ref{hcv2}).
Nevertheless, in \cite{flsz} the chirality mixing term was excluded of the EFT of the CS model given that
it gives zero contribution to first order in perturbation theory with parameter
$g$. The EFT was constructed by analytically extending the theory to larger values
of $g$ given that the $\winf$ symmetry determined the terms in the theory and there 
is one free parameter. Furthermore, the EFT defined with this procedure has successfully 
predicted the existence of Benjamin-Ono density waves at the operator level \cite{bz},
in agreement with the first quantized prediction of \cite{aw,abw}, and determined the dynamic structure factor \cite{dsf}
in agreement with the first quantized determination \cite{pustilink} in the overlapping 
regions of validity.
In addition, in \cite{boze3} we have considered this attractive interaction (in the originally repulsive regime of
$g > 0$) between opposite moving currents showing that it dominates the short distance behavior 
and proposing it as responsible for the destabilization of the ground state. 
A short range component in coordinate space arises due to the quantum regularization of this interaction. 
Although this term appears in the EFT hamiltonian derived with the straightforward procedure \cite{polch},
it allows for higher energy process than the ones considered by the other terms of (\ref{series}).
In fact, the first and second (choral and antichiral) terms in the second line of (\ref{hcv2})
involve low energy process, such that the exchanged momenta $p$ satisfy $0 \leq p < \sqrt{N} \Delta p$,
where $\Delta p = \pi/L = 1/(2R)$ is the quantum of momentum. The third term (mixed chiralities) in  (\ref{hcv2})
involves backward scattering processes with momenta $p$ that are kinematically bounded from below:
$ 2p_F \leq p$, implying $2N\Delta p \leq p $, which violates the dimensionless cutoff $O(\sqrt{N})$
adopted in the derivation of the EFT \cite{flsz}. This means that this term should be excluded from the
hamiltonian of the EFT, or, in another words, should be considered spurious. 

These results are confirmed by the bosonic form of the CS effective hamiltonian (\ref{hcsf}),
obtained after the Bogoliubov transformation (\ref{bogocs}) has been performed,
and the rapidity paremeter selected as explained before.  
Note that in (\ref{hcsf}) the effective interaction in the $1/N^2 $ is attractive,
with coupling $-g/\xi^2 $, where one factor $1/\xi$ is a consequence of the 
different normalization of the Heisenberg algebra for fermionic and
bosonic realizations of the $\winf$ algebra.
The extra $1/\xi$ factor cancels out when considering the global normalization energy scale,
leaving a $\xi$ independent bosonic energy.
\barr
{\cal H}_{CS} &=& E_{CS}\
\left\{\left[\frac{\sqrt{\xi}}{4}\,\w_0^0
+\frac{1}{N}\,\w_0^1+
\frac{1}{N^2}\left(\frac{1}{\sqrt{\xi}}\,\w_0^2
-\frac{\sqrt{\xi}}{12}\,\w_0^0 \right.\right.
\right.\nl
&&-\ \left.\left.\left.
\frac{g}{2\xi^2}\,\sum_{\ell=1}^\infty
\,\ell~\w_{-\ell}^0\,\w_\ell^0\right)
\right]+\left(\,W~\leftrightarrow~{\overline W}\,\right)
\right\}~~~.
\label{hcsf}
\earr
Ultimately, the disappearance of the terms that mix the R and L chiral sectors in the low energy EFT is expected
as the Fermi momentum provides a large energy scale. For the same reason, a chiral factorization 
is consistent with the absence of an interaction energy scale and, therefore, conformal invariance.

\section{Conclusions}

The properties of the fermionic CFTs we have discussed here are ultimately a consequence of the 
topology of the Fermi sea: a segment ending in two Fermi points, labeled as L and R. Moreover,
this structure is symmetric under the parity exchange  L $\leftrightarrow$ R. The marginal deformations 
considered in this paper do not change the topology of it, impliying that the EFT shound have a L-R Hilbert space structure 
isomorphic to that of a free theory. The map between these structures is given by a Bogoliubov transformation,
that acts on the L and R coordinate space as a Lorentz boost. In the language of \cite{wentop}, the
deformations considered here provide a meaningful parametrization of the conformal manifold
of the $(c,{\overline c})=(1,1)$ CFTs.

Regarding the Calogero-Sutherland EFT, our results provide further evidence that it may be thought of as a simple non-linear 
deformation of
the free Weyl CFT (the EFT of the Tomonaga-Luttinger model). A recent determination of the Dynamic Structure 
Factor gives further support to this conclusion \cite{dsf}. Moreover, the key role in establishing the
low energy structure of the EFT played by the effective short distance
interaction proposed in \cite{boze3} has been confirmed by the findings of this work. 

\section{Appendix: detailed calculations of the Bogoliubov transformation for the Calogero-Sutherland EFT }

Here we detail the calculations of the application of the Bogoliubov transformation (\ref{bogocs}) to the 
most relevant terms of the CS hamiltonian (\ref{series}). We employ the same notation as in section 4.
with the identifications $V^0 \leftrightarrow J$, $V^1 \leftrightarrow J^{2}/2$ and $V^2 \leftrightarrow J^{3}/3$.
As in section 4, we consider here the hamiltonian densities rather than the hamiltonian themselves
whenever convenient:
To order $1/N$, we have:
\barr
H_{(1)} &=& \left(1+\frac{g}{2}\right) \left( \frac{J^{2}}{2} +
\frac{{\overline J}^{2}}{2}\right)+\frac{g}{2} J {\overline J} \nl
&=&
\left[(1+\frac{g}{2})\cosh (2\beta) -\frac{g}{2}\sinh(2\beta) \right] \left(\frac{{\hat J}^{2}}{2} 
+\frac{{\hat {\overline J}}^{2}}{2}\right)\nl
&+&
\left[\frac{g}{2}\cosh (2\beta) -(1+\frac{g}{2})\sinh(2\beta) \right] {\hat J} {\hat {\overline J}}\ .
\label{ap1}
\earr
The condition that makes the mixed L-R term zero in (\ref{ap2}) is 
\beq
\tanh(2\beta) = \frac{g}{g+2}\ ,
\label{solbogcs}
\eeq
which implies $\exp(2\beta) = \sqrt{1+g}$. Notice that for small $g$, $ \sqrt{1+g} \simeq \xi$. 
Plugging this condition back into (\ref{ap1}), we obtain:
\beq
H_{(1)} = \sqrt{1+g} \left(\frac{{\hat J}^{2}}{2} +
\frac{{\hat {\overline J}}^{2}}{2}\right)\ .
\label{h1afbo}
\eeq
Now we consider the order $1/N^2$ in (\ref{series}), excluding the trivial terms 
$J+{\overline J} \leftrightarrow {\rm e}^{-\beta}\left( {\hat J}+{\hat {\overline J}} \right) $ 
and the momentum dependent interaction terms that are discussed separately in section 5:
\barr
H^{'}_{(2)} &=&
\left(1+\frac{g}{4}\right)\left(\frac{J^{3}}{3} +\frac{{\overline J}^{3}}{3} \right)
+\ \frac{g}{2} \left( \frac{J^{2}}{2} {\overline J}+ J \frac{{\overline J}^{2}}{2} \right) \nl
&=&
{\rm e}^{-\beta} \left[(1+\frac{g}{4})\left( \cosh (2\beta) + \frac{1}{2} \sinh(2\beta) \right) 
-\frac{3g}{8}\sinh(2\beta)  
\right] 
\left( \frac{{\hat J}^{3}}{3} + \frac{{\hat{\overline J}}^{3}}{3} \right) \nl
&+& 
{\rm e}^{-\beta} \left[ g\cosh(2\beta) -(2+g) \sinh(2\beta) \right]
\left( \frac{{\hat J}^{2}}{2} {\hat {\overline J}}+ {\hat J} \frac{{\hat {\overline J}}^{2}}{2} \right)\ .
\label{ap2}
\earr
The mixed terms vanish again with the condition (\ref{solbogcs}). We not do have an {\it a priori} argument
for this as to now. Plugging in back (\ref{solbogcs}) in (\ref{ap2}), we obtain:
\beq
H^{'}_{(2)} = (1+g)^{1/4} \left(\frac{{\hat J}^{3}}{3} +
\frac{{\hat {\overline J}}^{3}}{3}\right)\ .
\label{h2afbo}
\eeq
Expressions (\ref{h1afbo}) and (\ref{h2afbo}) agree with the non-interacting pieces of (\ref{hcsf}), 
once one realizes that $W^0 \leftrightarrow \sqrt{\xi} {\hat J}$, $W^1 \leftrightarrow {\hat J}^{2}/(2 \xi)$ 
and $W^2 \leftrightarrow {\hat J}^{3}/(3 \xi)$. The normalizations of the Sugawara constructions arise from
(\ref{kmalg1}).

Now we consider the three terms in the second line of (\ref{hcv2}), and apply the Bogoliubov transformation  
(\ref{bogocs}). We obtain:
\barr
H^{''}_{(2)} &=& 
-\ \frac{g}{4}\sum_{\ell=-\infty}^{\infty} |\ell |
\left( \v^0_{\ell}\,\v^0_{-\ell}+
\vb^0_{-\ell} \,\vb^0_{\ell}+
2\,\v^0_{\ell} \,\vb^0_{\ell}\right)\nl
&=&
-\ \frac{g}{4\xi} \sum_{\ell=-\infty}^{\infty}\ |\ell |\
\left( \w^0_{\ell}\,\w^0_{-\ell}+
\wb^0_{-\ell} \,\wb^0_{\ell} \right) \nl
&&-\ \frac{g}{2\xi}
\sum_{\ell=-\infty}^{\infty}\  
|\ell |\ \w^0_{\ell} \,\wb^0_{\ell} ~~~.
\label{ap3}
\earr

%
\def\NP{{\it Nucl. Phys.\ }}
\def\PRL{{\it Phys. Rev. Lett.\ }}
\def\PL{{\it Phys. Lett.\ }}
\def\PR{{\it Phys. Rev.\ }}
\def\IJMP{{\it Int. J. Mod. Phys.\ }}
\def\MPL{{\it Mod. Phys. Lett.\ }}

\end{document}